\documentclass{article}%
\usepackage{amsfonts}
\usepackage{amsmath}
\usepackage{amssymb}
\usepackage[latin1]{inputenc}
\usepackage[english]{babel}
\usepackage{graphicx}
\usepackage{placeins}
\usepackage[affil-it]{authblk}
\usepackage{booktabs}
\usepackage{dcolumn}
\usepackage{array}
\usepackage{longtable}%
\providecommand{\U}[1]{\protect\rule{.1in}{.1in}}
\newtheorem{postulate}{Postulate}[section]
\newtheorem{theorem}{Theorem}[section]
\newtheorem{corollary}{Corollary}[section]
\begin{document}

\title{\textbf{ Space-time singularities in Weyl manifolds }}
\author{I. P. Lobo\thanks{iarley.pereiralobo@icranet.org}
\and A. B. Barreto\thanks{adrianobraga@fisica.ufpb.br}
\and C. Romero\thanks{cromero@fisica.ufpb.br}}
\date{\today}
\maketitle

\begin{abstract}
We extend one of the Hawking-Penrose singularity theorems in general
relativity to the case of some scalar-tensor gravity theories in which the
scalar field has a geometrical character and space-time has the mathematical
structure of a Weyl integrable space-time (WIST). We adopt an invariant
formalism, so that the extended version of the theorem does not depend on a
particular frame.

\end{abstract}





\affil[1]{CAPES Foundation, Ministry of Education of Brazil, Bras\'ilia, Brazil and Sapienza Universit\`{a} di Roma - Dipartimento di Fisica\\
P.le Aldo Moro 5 - 00185 Rome - Italy}
\affil[2]{Universidade Federal da Para\'{i}ba, Departamento de F\'{i}sica,\\
C. Postal 5008, 58051-970 Jo\~ao Pessoa, PB, Brazil}



\section{Introduction}

Until the mid sixties it was argued by some cosmologists that the presence of
space-time singularities in general relativistic cosmological models were not
an essential property of the model, being, in fact, a consequence of the high
degree of symmetries of the distribution of matter assumed in these models
\cite{Khalatinikov}. Accordingly, it was believed that in more realistic
situations these singularities would disappear. However, this scenario was to
change drastically after a series of general mathematical results, namely, the
so-called Hawking-Penrose singularity theorems, were proved \cite{Penrose}. As
is widely known, these theorems use methods of global analysis to show that,
under the assumption of validity of general relativity and reasonable physical
behaviour of matter, space-time singularities are general phenomena which
occur in gravitational collapse and cosmology (such as the big bang)
irrespective of symmetry of the models.\newline

The investigation of space-time singularities has touched on some deep
philosophical issues and it is the view taken by many scientists that these
issues seem to call for a conceptual revision of general relativity that at
least take in account quantum mechanics. As far as cosmological singularities
are concerned it seems that the current view hold by most cosmologists is that
general relativity must break down at times less than the Planck time. In
fact, there has been a fair deal of work on classical and quantum new
approaches to gravity in which mechanisms that naturally prevent the formation
of a cosmological singularity are present. In fact, from the standpoint of
quantum mechanics the mere existence of a cosmological singularity would be in
contradiction with Heisenberg's uncertainty principle. In connection with this
point, much effort has been devoted to understand what could be called the
quantum structure of space-time. Among the numerous attempts to make progress
in this rather difficult issue, we would like to call attention for a recent
proposal which argues that the quantum structure of space-time may be even
related to the possibility of the space-time geometry possessing a
non-riemannian character \cite{novello3}. On the other hand, non-riemannian
geometries have appeared in physics mainly as a way to modify Einstein's
gravity while keeping the idea that the gravitational field is a manifestation
of the space-time geometry \cite{Goenner}. As is well known, one of the
earlier attempts in this direction was the Weyl's unified field theory
\cite{Weyl}.\newline

Concerning space-time singularities, we know that modified theories of gravity
call for a new mathematical treatment of the problem as the approach provided
by Hawking and Penrose is suitable only for general relativity. It is our
purpose here to pursue this question further in the light of some modified
gravity theories formulated in a particular kind of space-time geometry,
namely, the so-called Weyl integrable space-time (WIST) \cite{NH,
Weyl2}.\newline

The present article is organized as follows. In Section \ref{sec2}, we outline
the fundamental ideas of the geometry developed by Weyl, which underlies his
unified field theory. We then proceed to briefly review the main features of
Weyl's attempt to unify gravity and electrodynamics in a single geometric
framework. In Section \ref{sec3}, we consider the extension of Raychaudhuri
equation from a Riemannian setting to the case of a Weyl integrable
space-time. This extension is invariant under Weyl transformations. In Section
\ref{sec4}, we work out a generalization of one of the Hawking-Penrose
singularity theorems. Some applications of the formalism are given in Section
\ref{sec5}. These correspond to cosmological solutions of scalar-tensor
theories framed in a Weylian space-time. We conclude with some remarks in
Section \ref{sec6}.


\section{Weyl geometry\label{sec2}}

Weyl geometry arises from the weakening of one of the postulates of the
Riemannian geometry, similarly as non-Euclidean geometry was conveived after
the fifth postulate of Euclidean geometry was relaxed. The postulate we are
referring to is the so-called \textit{the Riemannian compatibility condition},
which states the following:

\begin{postulate}
Let $M$ be a differentiable manifold endowed with an affine connection
$\nabla$ and a metric tensor $g$. For any vector fields $U,V,$ $W\in T(M)$, it
is required that
\begin{equation}
V[g(U,W)]=g(\nabla_{V}U,W)+g(U,\nabla_{V}W).\label{Riemannian compatibility}%
\end{equation}

\end{postulate}

As is well known, this condition is equivalent to demand that the covariant
derivative of the metric vanish, thereby implying that the length of a vector
remain unaltered by parallel transport.\newline

On the other hand, in Riemannian geometry, it will also be assumed that the
connection $\nabla$ be \textit{torsionless} (or \textit{symmetric}), i.e.,
that for any $U,V\in T(M)$ the following condition holds:
\begin{equation}
\nabla_{V}U-\nabla_{U}V=[V,U].\label{torsionless}%
\end{equation}
From these two postulates we are led to the important Levi-Civita theorem,
which states that the affine connection is entirely determined from the metric
\cite{do Carmo}. In a certain sense, this theorem characterizes a Riemannian
manifold.\newline

In 1918 the mathematician Hermann Weyl generalized the geometry of Riemann by
introducing the possibility of change in the length of a vector through
parallel transport. To implement this idea Weyl conceived the following new
compatibility condition:

\begin{postulate}
[Weyl]Let $M$ be a differentiable manifold endowed with an affine connection
$\nabla$, a metric tensor $g$ and a one-form field $\sigma$, called a Weyl
field. It is said that $\nabla$ is compatible (W-compatible) with $g$ if for
any vector fields $U,V,$ $W\in T(M)$, we have
\begin{equation}
V[g(U,W)]=g(\nabla_{V}U,W)+g(U,\nabla_{V}W) + \sigma
(V)g(U,W).\label{W-compatible}%
\end{equation}

\end{postulate}

Clearly, this represents a generalization of the Riemannian compatibility
condition. Naturally, if the one-form $\sigma$ vanishes, we reobtain
(\ref{Riemannian compatibility}). In this way, we have a generalized version
of the Levi-Civita theorem given by the following proposition (see
Ref.:\cite{Gomez} for a proof):

\begin{theorem}
[Levi-Civita extended]Let $M$ be a differentiable manifold endowed with a
metric $g$ and a differentiable one-form field $\sigma$ defined on $M$, then
there exists one and only one affine connection $\nabla$ such that: $i)$
$\nabla$ is torsionless; $ii)$ $\nabla$ is W-compatible with $g$.
\end{theorem}

It follows that in a coordinate basis $\{x^{a}\}$ one can express the
components of the affine connection completely in terms of the components of
$g$ and $\sigma$:%
\begin{equation}
\Gamma_{bc}^{a}=\{_{bc}^{a}\}-\frac{1}{2}g^{ad}[g_{db}\sigma_{c}+g_{dc}%
\sigma_{b}-g_{bc}\sigma_{d}] \label{Weylconnection}%
\end{equation}
where $\{_{bc}^{a}\}=$ $\frac{1}{2}g^{ad}[g_{db,c}+g_{dc,b}-g_{bc,d}]$ denotes
the Christoffel symbols of second kind. The next proposition gives a helpful
insight on the geometrical meaning of the Weyl parallel transport.

\begin{corollary}
Let $M$ be a differentiable manifold with an affine connection $\nabla$, a
metric $g$ and a Weyl field of one-forms $\sigma$. If $\nabla$ is
W-compatible, then for any smooth curve $\alpha=\alpha(\lambda)$ and any pair
of two parallel vector fields $V$ and $U$ along $\alpha,$ we have
\begin{equation}
\frac{d}{d\lambda}g(V,U)=\sigma(\frac{d}{d\lambda})g(V,U)
\label{covariantderivative}%
\end{equation}
where $\frac{d}{d\lambda}$ denotes the vector tangent to $\alpha$.
\end{corollary}

By integrating the above equation along the curve $\alpha$ from a point
$p_{0}=\alpha(\lambda_{0}),$ we get
\begin{equation}
g(V(\lambda),U(\lambda))=g(V(\lambda_{0}),U(\lambda_{0}))e^{\int_{\lambda_{0}%
}^{\lambda}\sigma(\frac{d}{d\rho})d\rho}. \label{integral}%
\end{equation}
Let us set $U=V$ and denote by $L(\lambda)$ the length of the vector
$V(\lambda)$ at an arbitrary point $p=\alpha(\lambda)$ of the curve. It is
easy to check that in a local coordinate system $\left\{  x^{a}\right\}  $ the
equation (\ref{covariantderivative}) reduces to
\begin{equation}
\frac{dL}{d\lambda}=\frac{\sigma_{a}}{2}\frac{dx^{a}}{d\lambda}L
\end{equation}
Consider now the set of all closed curves $\alpha:[a,b]\in R\rightarrow M$,
i.e, with $\alpha(a)=\alpha(b).$ The equation
\begin{equation}
g(V(b),U(b))=g(V(a),U(a))e^{\int_{a}^{b}\sigma(\frac{d}{d\lambda})d\lambda}
\label{holonomy}%
\end{equation}
defines a holonomy group. If we want the elements of this group to correspond
to an isometry, then we must require that%
\begin{equation}
\oint\sigma\left( \frac{d}{d\lambda}\right) d\lambda=0
\end{equation}
for any loop. From Stokes' theorem it follows that $\sigma$ must be an exact
form, that is, there exists a scalar function $\phi$, such that $\sigma=d\phi
$. In this case we have a \textit{Weyl integrable manifold}.\newline

Weyl manifolds are completely characterized by the set $(M,g,\sigma)$, which
will be called a \textit{Weyl frame}. Let us remark that the Weyl
compatibility condition (\ref{covariantderivative}) remains unchanged when we
go to another Weyl frame $(M,\overline{g},\overline{\sigma})$ by carrying out
the following simultaneous transformations in $g$ and $\sigma$:%
\begin{align}
\overline{g}=e^{f}g\text{ ,}\label{conformal}\\
\nonumber\\
\overline{\sigma}=\sigma+df\text{ ,}\label{gauge}%
\end{align}
where $f$ is an arbitrary scalar function defined on $M$. The conformal map
(\ref{conformal}) and the \textit{gauge }transformation (\ref{gauge}) define
classes of equivalences in the set of Weyl frames. It is important to mention
that the discovery that the compatibility condition (\ref{covariantderivative}%
) is invariant under this group of transformations was essential to Weyl's
attempt at unifying gravity and electromagnetism, extending the concept of
space-time to that of a collection of manifolds equipped with a conformal
structure, leading to the notion that space-time might be viewed as a class
$[g]$ of conformally equivalent Lorentzian metrics \cite{Weyl}.


\section{Weyl integrable space-time\label{sec3}}

We now consider the particular case of a Weyl integrable space-time (WIST),
where $\sigma=d\phi$. As already mentioned, the set $(M,g,\phi)$ consisting of
a differentiable manifold $M$ endowed with a metric $g$ and a Weyl scalar
field $\phi$ will be referred to as a \textit{Weyl frame}. In this case
(\ref{gauge}) becomes%
\begin{equation}
\overline{\phi}=\phi+f.
\end{equation}
If we set $f=-\phi$ in the above equation, we get $\overline{\phi}=0$. We
refer to the set $(M,\overline{g}=e^{-f}g,\overline{\phi}=0)$ as the
\textit{Riemann frame}, because in this frame the manifold becomes Riemannian.
On the other hand, it can be easily verified that the equation
(\ref{Weylconnection}) follows directly from $\nabla_{\alpha}\overline{g}%
_{\mu\nu}=0$. This result has interesting and useful consequences. In fact,
the metric $\gamma=e^{-\phi}g$, defined for any frame $(M,g,\phi)$, is
invariant under the Weyl transformations (\ref{conformal}) and (\ref{gauge})
any geometric quantity built exclusively with $\gamma$ is invariant. More
generally, geometric objects such as the components of the curvature tensor
$R_{\beta\mu\nu}^{\alpha}$ , the components of the Ricci tensor $R_{\mu\nu}$ ,
the scalar $e^{\phi}R$ are invariant under the Weyl transformations
(\ref{conformal}) and (\ref{gauge}).\newline

It is important to note that because the Weyl transformations (\ref{conformal}%
) and (\ref{gauge}) define an equivalence relation between frames $(M,g,\phi)$
it seems more appropriate to look into the equivalence class of such frames
rather than on a particular frame. In other words, a Weyl manifold should be
regarded as a frame $(M,g,\phi)$ that is only defined \textquotedblleft up to
a Weyl transformation\textquotedblright. In this way, when dealing with a
certain Weyl manifold we chose a particular frame in the equivalence class,
and consider that only geometric entities defined in that frame which are
invariant are of interest, since they can be regarded as representative of the
whole class. From this point of view, it is more natural to redefine some
Riemannian concepts to meet the requirements of invariance. This procedure is
analogous to the one adopted in \textit{conformal geometry}, a branch of
geometry, in which the geometric objects of interest are those invariant under
conformal transformation, such as, say, the angle between two directions
\cite{conformal}. Accordingly, one should naturally generalize the definition
of all invariant integrals when dealing with the integration of exterior
forms. For example, the Riemannian $q$-dimensional volume form defined as
$\Omega=\sqrt{-g}dx^{1}\wedge...\wedge dx^{q}$ is not invariant under Weyl
transformations, hence it should be replaced by $\Omega=\sqrt{-\gamma
}e^{-\frac{q}{2}\phi}dx^{1}\wedge...\wedge dx^{q}$, and so on\footnote{Note
that $g$ in the expression $\sqrt{-g}$ denotes the determinant of $g_{\mu\nu}%
$.}. Likewise, in a Weyl integrable manifold it is more natural to define the
concept of \textquotedblleft length of a curve\textquotedblright in an
invariant way. It follows that our familiar notion of proper time as the arc
length of worldlines in four-dimensional Lorentzian space-time should be
modified. Because of this, we shall redefine the proper time $\Delta\tau$
measured by a clock moving along a parametrized timelike curve $x^{\mu}%
=x^{\mu}(\lambda)$ between $x^{\mu}(a)$ and $x^{\mu}(b)$ , in such a way, that
$\Delta\tau$ is the same in all frames. This suggests the following
definition:
\begin{equation}
\Delta\tau= \int_{a}^{b}\left(  \overline{g}_{\mu\nu}\frac{dx^{\mu}}{d\lambda
}\frac{dx^{\nu}}{d\lambda}\right)  ^{\frac{1}{2}}d\lambda= \int_{a}
^{b}e^{-\frac{\phi}{2}}\left(  g_{\mu\nu}\frac{dx^{\mu}}{d\lambda}%
\frac{dx^{\nu}}{d\lambda}\right)  ^{\frac{1}{2}}d\lambda.\label{propertime}%
\end{equation}
It must be noted that the above expression can be also obtained from the
special relativistic definition of proper time if we adopt the prescription
$\eta_{\mu\nu}\rightarrow e^{-\phi}g_{\mu\nu}.$ It is clear that the
right-hand side of this equation is invariant under Weyl transformations and
that, in the Riemann frame, it reduces to the definition of proper time in
general relativity. We, therefore, take $\Delta\tau$, as given above, as the
extension to an arbitrary Weyl frame of general relativistic clock hypothesis,
i.e., the assumption that $\Delta\tau$ measures the proper time measured by a
clock attached to the particle.\newline

It is now easy to see that the extremization of the functional
(\ref{propertime}) leads to the equations
\begin{equation}
\frac{d^{2}x^{\mu}}{d\lambda^{2}}+\left( \left\{  _{\alpha\beta}^{\mu}\right\}
-\frac{1}{2}g^{\mu\nu}(g_{\alpha\nu}\phi_{,\beta}+g_{\beta\nu}\phi_{,\alpha
}-g_{\alpha\beta}\phi\,_{,\nu})\right) \frac{dx^{\alpha}}{d\lambda}%
\frac{dx^{\beta}}{d\lambda}=0, \label{geodesics}%
\end{equation}
where $\left\{  _{\alpha\beta}^{\mu}\right\}  $ designates the Christoffel
symbols calculated with $g_{\mu\nu}$. Recall that in the derivation of the
above equations the parameter $\lambda$ must be chosen such that%
\begin{equation}
e^{-\phi}g_{\alpha\beta}\frac{dx^{\alpha}}{d\lambda}\frac{dx^{\beta}}%
{d\lambda}=K=const \label{constant}%
\end{equation}
along the curve, which, up to an affine transformation, permits us to identify
$\lambda$ with the proper time $\tau$. Surely, these equations are exactly
those that yield the affine geodesics in a Weyl integrable space-time, since
they can be rewritten as
\begin{equation}
\frac{d^{2}x^{\mu}}{d\tau^{2}}+\Gamma_{\alpha\beta}^{\mu}\frac{dx^{\alpha}%
}{d\tau}\frac{dx^{\beta}}{d\tau}=0, \label{Weylgeodesics}%
\end{equation}
where $\Gamma_{\alpha\beta}^{\mu}=\left\{  _{\alpha\beta}^{\mu}\right\}
-\frac{1}{2}g^{\mu\nu}(g_{\alpha\nu}\phi_{,\beta}+g_{\beta\nu}\phi_{,\alpha
}-g_{\alpha\beta}\phi\,_{,\nu})$, according to (\ref{Weylconnection}), are
identified with the components of the Weyl connection. Thus the extension of
the geodesic postulate by requiring that the functional (\ref{propertime}) be
an extremum is equivalent to assuming that the particle motion must follow
affine geodesics defined by the Weyl connection $\Gamma_{\alpha\beta}^{\mu}$.
Let us note that, as a consequence of the Weyl compatibility condition
(\ref{W-compatible}) between the connection and the metric, (\ref{constant})
holds automatically along any affine geodesic determined by
(\ref{Weylgeodesics}). Since both the connection components $\Gamma
_{\alpha\beta}^{\mu}$ and the proper time $\tau$ are invariant when we switch
from one Weyl frame to the other, the equations (\ref{Weylgeodesics}) are
invariant under Weyl transformations.\newline

As is well known, the geodesic postulate is not only concerned with the motion
of particles, but also determines the propagation of light rays in space-time.
On the other hand, since the path of light rays are null curves, one cannot
use the proper time as a parameter to describe these curves. Thus light rays
are supposed to follow null affine geodesics, which cannot be defined in terms
of the functional (\ref{propertime}), but, instead, they must be characterized
by their behaviour with respect to parallel transport. We naturally extend
this postulate by simply assuming that light rays follow Weyl null affine
geodesics.

\section{The Raychaudhuri equation\label{sec4}}

The Raychaudhuri equation played a fundamental role in the derivation of the
Hawking-Penrose singularity theorems. In the derivation, however, it is
assumed right from the beginning that the geometry underlying the space-time
is Riemannian. In this section, we investigate the extension of this equation
to the case of a Weyl integrable space-time.\newline

Let us first remark that the extension of the geodesic postulate to WIST
assumes that the particle motion must follow Weyl time-like geodesics. In the
following we shall consider a smooth congruence $\Gamma$ of time-like
geodesics corresponding to the worldlines of a class of observers,
parametrized by the invariant proper time $\tau$ defined in (\ref{propertime}%
). Hence, the tangents to the congruence generate a tangent vector field $V$
normalized to unit length. In order to keep the formalism invariant under Weyl
transformations we shall choose the affine parameter of the congruence as the
Weyl invariant arc length, i.e., we normalize $V$ with respect to the
invariant metric $\gamma_{\mu\nu}=e^{-\phi}g_{\mu\nu}$.
\begin{equation}
\gamma(V(\tau),V(\tau))=1. \label{normalization}%
\end{equation}
Therefore, in a local coordinate system, a geodesic curve described by
$x^{\mu}(\tau)$ satisfies $V^{\mu}\nabla_{\mu}V^{\alpha}=0$, where $V^{\alpha
}=\frac{dx^{\alpha}}{d\tau}$ and we have the affine geodesics in a Weyl
integrable space-time shown in (\ref{Weylgeodesics}). Furthermore, once we
have normalized the tangent vector field with the invariant metric
$\gamma_{\mu\nu}$, we obtain
\begin{equation}
V_{\mu}\nabla_{\alpha}V^{\mu}=0=V^{\nu}\nabla_{\alpha}V_{\nu},
\label{invarcond}%
\end{equation}
since $\nabla_{\alpha}\gamma_{\mu\nu}=0$.\newline

Let us now consider, at some point $p$ of $M$, the hypersurface $\Sigma$
orthogonal to the vector field $V$. As in the standard procedure, we define
the operator of projection $\Pi$ onto the hypersurface $\Sigma$ as
\begin{equation}
\Pi_{\mu\nu}=\gamma_{\mu\nu}-V_{\mu}V_{\nu}. \label{projectensor}%
\end{equation}
As is well known, $\Pi_{\mu\nu}$ represents to the first fundamental form of
the hypersurface induced by the metric $\gamma$ and its role is to project any
vector of $T_{p}M$ at $p$ onto $T_{p}\Sigma$, the tangent space to the
submanifold $\Sigma$\footnote{Note that we are using the invariant metric
$\gamma$ in order to ensure a invariant projection tensor. Accordingly, the
operation of raising and lowering tensorial indices must be always carried out
with $\gamma$. This guarantees that the duality between covariant and
contravariant vectors is not modified by Weyl transformations.}.\newline

We proceed with the derivation of the Raychaudhuri equation in this new
setting. We first need to consider a smooth one-parameter subfamily
$\alpha_{s}(\tau)$ of geodesics in the congruence of $V$ and then define a
deviation vector $\eta$ that represents an infinitesimal spatial displacement
from a given geodesic $\alpha_{o}(\tau)$ to a neighboring geodesic in this
subfamily. Once we have been given $\eta$ in $\Sigma$, we define the vector
field $\eta$ along this subfamily by Lie dragging it along $V$, that is, by
requiring that
\begin{equation}
£_{V}(\eta)=0.
\end{equation}
From the definition of Lie derivative and the fact that the connection
$\nabla$ is assumed to be torsionless we are led to the following equation
\begin{equation}
V^{\mu}\nabla_{\mu}\eta^{\alpha}=\eta^{\mu}\nabla_{\mu}V^{\alpha},
\label{equiv}%
\end{equation}
which shows how the deviation vector changes along the congruence. An
important role played in the investigation of the behaviour of neighbouring
geodesics as we go along the congruence $\Gamma$ is played by the so-called
deformation tensor $Q_{\mu\nu}$, defined as
\begin{equation}
Q_{\mu\nu}=\nabla_{\nu}V_{\mu}. \label{deformationtensor}%
\end{equation}
In terms of $Q_{\mu\nu}$ we can rewrite (\ref{equiv}) as
\begin{equation}
V^{\mu}\nabla_{\mu}\eta^{\alpha}\equiv Q_{\;\mu}^{\alpha}\eta^{\mu},
\end{equation}
which clearly means that the deformation tensor measures the failure of
$\eta^{\mu}$ to be parallelly transported \cite{wald}. Furthermore, it is easy
to see that this tensor is purely spatial, since $Q_{\mu\nu}V^{\mu}%
=0=Q_{\mu\nu}V^{\nu}$ . Thus, $Q_{\mu\nu}$ is a tensor defined in the subspace
of the tangent space perpendicular to $V$. Finally, in order to have a
physical interpretation of some kinematical aspects of the congruence $\Gamma$
, we can decompose $Q_{\mu\nu}$ in its irreducible parts:
\begin{equation}
Q_{\mu\nu}\equiv\frac{1}{3}\Theta\Pi_{\mu\nu}+\sigma_{\mu\nu}+\omega_{\mu\nu},
\label{Birreducible}%
\end{equation}
where the parameters $\Theta$, $\sigma_{\mu\nu}$, and $\omega_{\mu\nu}$ are
known, respectively, as the expansion, shear and vorticity of the congruence
(for instance, Ref.:\cite{wald}). From the above equation we have
\begin{align}
\Theta &  =\Pi^{\mu\nu}Q_{\mu\nu},\label{expansion}\\
\sigma_{\mu\nu}  &  =Q_{(\mu\nu)}-\frac{1}{3}\Theta\Pi_{\mu\nu},
\label{shear}\\
\omega_{\mu\nu}  &  =Q_{[\mu\nu]}. \label{twist}%
\end{align}
It is easy to see that in the case where the congruence is locally orthogonal
to the hypersurface $\Sigma$, we have $\omega_{\mu\nu}=0$.\newline

We are now particularly interested in the behaviour of $\Theta$, which
measures the expansion of the congruence and can tell us about the existence
of conjugated points. Thus, in order to obtain the Raychaudhuri equation, we
need to know the rate of change of $\Theta$ along the congruence. Therefore,
since, by definition, $\Theta=\nabla_{\alpha}V^{\alpha}$, we must compute
$V^{\mu}\nabla_{\mu}\Theta=V^{\mu}\nabla_{\mu}[\nabla_{\alpha}V^{\alpha}].$ On
the other hand, from the definition of the curvature tensor, we have
\begin{equation}
\nabla_{\beta}\nabla_{\nu}V^{\mu}-\nabla_{\nu}\nabla_{\beta}V^{\mu
}=R_{\;\lambda\nu\beta}^{\mu}V^{\lambda}%
\end{equation}
where,
\begin{equation}
R_{\;\;\lambda\nu\beta}^{\mu}=\partial_{\beta}\Gamma_{\lambda\nu}^{\mu
}-\partial_{\nu}\Gamma_{\lambda\beta}^{\mu}+\Gamma_{\lambda\nu}^{\rho}%
\Gamma_{\rho\beta}^{\mu}-\Gamma_{\lambda\beta}^{\rho}\Gamma_{\rho\nu}^{\mu},
\end{equation}
one can readily obtain
\begin{equation}
V^{\mu}\nabla_{\mu}\Theta=-\frac{1}{3}\Theta^{2}-2(\sigma^{2}-\omega
^{2})+R_{\mu\nu}V^{\mu}V^{\nu}, \label{Ray}%
\end{equation}
where are denoting $\sigma^{2}=\sigma_{\mu\nu}\sigma^{\mu\nu}$, $\omega
^{2}=\omega_{\mu\nu}\omega^{\mu\nu}$, and the term $R_{\mu\nu}V^{\mu}V^{\nu}$
is usually referred to as the Raychaudhuri scalar. Setting $\dot{\Theta
}=V^{\mu}\nabla_{\mu}\Theta$, we can write (\ref{Ray}) in the following form,
known as the Raychaudhuri equation:
\begin{equation}
\dot{\Theta}+\frac{1}{3}\Theta^{2}+2(\sigma^{2}-\omega^{2})=R_{\mu\nu}V^{\mu
}V^{\nu}. \label{Ray1}%
\end{equation}
It is worth noting that (\ref{Ray1}) has the same form as in the case of a
Riemannian space-time, although it must be recalled that the Ricci tensor is
built with the Weylian connection. In fact, this is not surprising as we have
redefined the proper time in an invariant way, using the invariant metric
$\gamma_{\mu\nu}=e^{-\phi}g_{\mu\nu}$. In the next section, we shall analyse
the conditions that lead to singularities in the space-time.

\section{Extending the Singularity Theorem\label{sec5}}

Because of the form of the Raychaudhuri equation takes in a Weyl integrable
space-time, the description of conjugate points is the same as in Riemannian
geometry. Thus, we have the following statement: if $\theta_{0}=\theta
(\tau_{0})<0$ for some $\tau=\tau_{0}$, $R_{\mu\nu}V^{\mu}V^{\nu}\leq0$, then
in a finite invariant proper time $\tau\leq3/|\theta_{0}|$, the congruence
will develop a conjugate point $\theta\rightarrow-\infty$. As a matter of
fact, the extension of any theorem from Riemannian geometry to a Weyl
integrable can be trivially carried out by simpling considering the invariant
metric $\gamma=e^{-\phi}g$. For instance, as we have already mentioned the
extremization of the functional (\ref{propertime}) leads directly to the
Weylian geodesics. On the other hand, results coming from differential
topology and concerning the causal structure of space-time that are valid in a
Riemannian space-time are also valid in a Weyl integrable space-time since
conformal transformations do not affect the light-cone structure nor the
manifold orientability. Indeed, the proof of one of the most important results
we are going to enunciate now, namely, the generalized Jacobi theorem,
proceeds along the same lines of reasoning employed in the Riemannian case,
where we merely replace $g$ by the invariant metric $\gamma=e^{-\phi}g$
\cite{wald}.

\begin{theorem}
[Jacobi]Let $\gamma:[0,1]\rightarrow M$ a differential time-like curve
connecting points $p$, $q\in M$. Then a necessary and sufficient condition for
$\gamma$ to locally maximize the invariant arc length between $p$ and $q$ is
that $\gamma$ be a Weyl geodesic without conjugate points between $p$ and $q$.
\end{theorem}

Note the relation between the Raychaudhuri scalar $\Theta$ and the extrinsic
curvature of the submanifold orthogonal to geodesic congruence, which is
represented by the mixed tensor $Q_{\;\nu}^{\mu}=\Pi_{\ \alpha}^{\mu}%
\Pi_{\ \nu}^{\beta}\nabla_{\beta}V^{\alpha}$. For the case of a congruence of
geodesics orthogonal to hypersurface $\Sigma$ and parameterized by the
invariant arc length we have $Q_{\;\nu}^{\mu}=\nabla_{\nu}V^{\mu}$. Thus, the
trace of $Q_{\;\nu}^{\mu}$ is equal to $\Theta$, i.e., $Q=Q_{\;\alpha}%
^{\alpha}=\Theta$. Now we are ready to prove the following proposition:

\begin{theorem}
Let $(M,g,\phi)$ be a globally hyperbolic Weylian space-time with $R_{\mu\nu
}V^{\mu}V^{\nu}\leq0$ for all time-like vectors $V$. Suppose there exists a
space-like Cauchy surface $\Sigma$ such that the trace of its extrinsic
curvature (for the orthonormal congruence of past directed geodesics)
satisfies $B\leq C<0$ over the whole surface, $B$ and $C$ being
constants.\newline Then, no past directed time-like curve coming from $\Sigma$
can have an invariant arc length greater than $3/|C|$. In particular, all past
directed time-like geodesics are incomplete.
\end{theorem}

\textit{Proof}: Let us prove this theorem by contradiction. Suppose there
exists a time-like curve $\lambda=\lambda(\tau)$ coming from $\Sigma$ whose
value of the invariant arc length $\tau$ at some point $p$ $\in\lambda$ is
greater than $3/|C|$. It is known that as the set of curves joining two points
in a globally hyperbolic manifold is compact, the invariant arc length
function must have a maximum value for a given curve. Then, this curve must be
a Weyl geodesic. Therefore, there must be a geodesic $\gamma$ (with invariant
arc length greater than $3/|C|$) joining $p$ to $\Sigma$. This means there are
no conjugated points between $p$ and $\Sigma$. But, from Raychaudhuri's
inequality, we know that $\gamma$ must have conjugated points between $p$ and
$\Sigma$, which is a contradiction. We then conclude that the original curve
$\lambda$ cannot exist. $\Box$\newline

In the case of general relativity where the space-time mathematical structure
is that of Riemannian geometry, which is a special case of Weyl geometry when
$\phi$ is a constant of motion, the geometric condition
\begin{equation}
R_{\mu\nu}V^{\mu}{V}^{\nu}\leq0 \label{condition}%
\end{equation}
is equivalent to the so-called strong energy condition
\begin{equation}
T_{\mu\nu}V^{\mu}{}^{\nu}-T/2\geq0, \label{energy}%
\end{equation}
and requiring that $Q<0$ is equivalent to assuming that in the course of the
cosmic evolution the Universe underwent an expansion period, which seems to be
a rather reasonable assumption. In view of the above, this leads to the
conclusion that the Universe, as modelled by GR, must necessarily have had a
beginning starting from a singular state. Let us now consider some different
scenarios offered by two alternative gravitational theories, namely, the
Novello-Oliveira-Salim-Elbaz's theory (N) \cite{NH,Novello4} and a recent
geometrical approach to scalar-tensor theory (GST), both inspired in the idea
of that space-time can be described by Weyl integrable geometry \cite{Almeida}.

\subsection{Novello's theory}

Novello's theory starts with the action
\begin{equation}
S=\int d^{4}x\sqrt{-g}\left\{  R-2\xi\phi^{,\mu}\phi_{,\mu}+e^{-2\phi}%
L_{m}\right\}  , \label{action wist}%
\end{equation}
where $R$ denotes the Weylian scalar curvature, $\phi$ represents the Weyl
scalar field, $\xi$ is a dimensionless parameter, while $L_{m}$ stands for the
Lagrangian of the matter fields. The form of $L_{m}$ is determined from the
corresponding Lagrangian in special relativity by replacing the ordinary
derivatives by covariant derivatives with respect to the Weyl connection. The
field equations are given by
\begin{equation}
R_{\mu\nu}-\frac{1}{2}g_{\mu\nu}=-\nabla_{\mu}\phi_{,\nu}+(2\xi-1)\phi_{,\mu}
\phi_{,\nu}-\xi\phi_{,\alpha}\phi^{,\alpha}g_{\mu\nu}-e^{-\phi}T_{\mu\nu},
\label{NH2}%
\end{equation}%
\begin{equation}
\nabla_{\alpha}\phi^{,\alpha}+2\phi_{,\alpha}\phi^{,\alpha}=-\frac{1}%
{2\lambda}e^{-\phi}T\ \ , \label{fi}%
\end{equation}
where we have set $\lambda=\frac{3}{2}-2\xi.$\newline

From the above equations it is not difficult to verify that we have
\begin{equation}
R_{\mu\nu}V^{\mu}V^{\nu}=-\frac{d^{2}\phi}{d\bar{s}^{2}}+(2\xi-1)\left(
\frac{d\phi}{d\bar{s}}\right)  ^{2}-e^{-\phi}\left(  T_{\mu\nu}V^{\mu}V^{\nu
}-\frac{1}{2}|V|^{2}T+\frac{1}{4\lambda}|V|^{2}T\right)  .
\end{equation}
Clearly, in this theory (\ref{condition}) does not imply the violation of the
strong energy condition (\ref{energy}), but rather we have the following
situation. Consider the conditions below:
\begin{align}
\frac{d^{2}\phi}{d\bar{s}^{2}}  &  \equiv\ddot{\phi}\geq0,\\
2\xi-1  &  \leq0,\\
T_{\mu\nu}V^{\mu}V^{\nu}-\frac{1}{2}|V|^{2}T+\frac{1}{4\lambda}|V|^{2}T  &
\geq0.
\end{align}
If any one of these conditions is violated, then the solution may correspond
to a nonsingular space-time. Consider, as an example, the vacuum solution of
the field equations (\ref{NH2}) and (\ref{fi}) obtained in Ref.:
\cite{Novello4}. By assuming homogeneity and isotropy we can write
\begin{equation}
ds^{2}=dt^{2}-a^{2}(t)\left(  \frac{dr^{2}}{1-\kappa r^{2}}+r^{2}d\Omega
^{2}\right)  , \label{FRW}%
\end{equation}%
\begin{equation}
\phi_{,\alpha}=\phi_{,0}\delta_{\alpha}^{0}\doteq\dot{\phi}\delta_{\alpha}%
^{0}. \label{vi}%
\end{equation}
This leads to the equations
\begin{align}
\dot{a}^{2}+\kappa+\frac{(4\xi-3)}{6}(\dot{\phi}a)^{2}  &  =0,\\
2a\ddot{a}+\dot{a}^{2}+\kappa-\frac{4\xi-3}{4}(\dot{\phi}a)^{2}  &  =0,\\
(a^{3}\dot{\phi})_{,0}  &  =0.
\end{align}
By integrating the last of these equations we get $\dot{\phi}=\zeta a^{-3}$,
where $\zeta$ is a constant. Now, if $4\xi-3>0$, it follows that
\begin{equation}
\dot{a}^{2}(t) = 1 - \left[  \frac{a_{0}}{a(t)}\right] ^{4},\label{scale}%
\end{equation}
with $(a_{0})^{4}=(\frac{4\xi-3}{12})\zeta^{2}$, thus implying that $a(t)\geq
a_{0}$. The model describes a non-singular bouncing universe, a scenario in
which after undergoing a period of contraction, dominated by the scalar field,
the universe reaches a minimum value, and then starts to expand, at an
inflationary rate, until the radiation dominates over the scalar field and the
scale factor begins to follow the standard cosmic evolution \cite{Novello4}.
Clearly, the non-singular behaviour of the model is a consequence of the fact
that the condition $\xi\leq1/2$ is violated. In order to display the bouncing
character of this model, it is convenient to solve (\ref{scale}) in terms of
the conformal time $\eta$ , defined as $d\eta=dt/a(t)$. It can be shown that
the scale factor and the scalar field are given, respectively, by
\begin{align*}
a(\eta)  &  =a_{0}\sqrt{\cosh2(\eta-\eta_{0})},\\
\phi(\eta)  &  =\frac{\zeta}{2a_{0}^{2}}\arccos\left[  \cosh2(\eta-\eta
_{0})\right]  ^{-1}.
\end{align*}

\subsection{Geometrical scalar-tensor theory}

The geometrical approach to scalar-tensor theory starts with the Brans-Dicke
action
\begin{equation}
S=\int d^{4}x\sqrt{-g}e^{-\phi}[R+\omega\phi^{,\alpha}\phi_{,\alpha}+k^{\ast
}e^{-\phi}L_{m}(e^{-\phi}g)], \label{action1}%
\end{equation}
where $\phi$ is assumed \textit{a priori} to be a geometrical field, i.e., an
intrinsic part of the space-time geometry, $\omega$ is a dimensionless
parameter and $k^{\ast}=\frac{8\pi}{c^{4}}$. By applying the Palatini
variational method, one obtains the Weyl integral compatibility condition
\cite{Almeida}
\begin{equation}
\nabla_{\alpha}g_{\mu\nu}=g_{\mu\nu}\phi_{,\alpha}. \label{theory06}%
\end{equation}
Naturally, the action (\ref{action1}) can be easily extended to accommodate a
scalar potential $\mathcal{V}(\phi)$ and to allow for a functional dependence
of $\omega$ on $\phi$, thus leading to \cite{NewPucheu}
\begin{equation}
S = \int d^{4}x\sqrt{-g}\{e^{-\phi}\left[  R+\omega(\phi)g^{\mu\nu}\phi_{,\mu
}\phi_{,\nu}\right] -\mathcal{V}(\phi) + \kappa^{\ast}e^{-2\phi}%
L_{m}\}.\label{action2}%
\end{equation}
The field equations obtained from (\ref{action1}) are given by
\begin{equation}
R_{\mu\nu}-\frac{1}{2}g_{\mu\nu}R =\omega(\phi)\left(  \frac{1}{2}g_{\mu\nu
}\phi_{,\alpha}\phi^{,\alpha}-\phi_{,\mu}\phi_{,\nu}\right)  -\frac{1}%
{2}e^{\phi}g_{\mu\nu}\mathcal{V}(\phi)-\kappa^{\ast}T_{\mu\nu},\label{FE1-GBD}%
\end{equation}
\begin{equation}
\Box\phi=-\left(  1+\frac{1}{2\omega}\frac{d\omega}{d\phi}\right)
\phi_{,\alpha}\phi^{,\alpha}-\frac{e^{\phi}}{\omega}\left(  \frac{1}{2}%
\frac{d\mathcal{V}}{d\phi}+\mathcal{V}\right)  , \label{FE2-GBD}%
\end{equation}
where $\Box$ denotes the d'Alembertian operator calculated with the Weyl
connection.\newline

By taking into account (\ref{FE1-GBD}) one easily obtains
\begin{equation}
R_{\mu\nu}V^{\mu}V^{\nu}\equiv-\kappa^{\ast}\left(  T_{\mu\nu}V^{\mu}V^{\nu
}-\frac{1}{2}|V|^{2}T\right)  -\omega(\phi)\dot{\phi}^{2}+|V|^{2}e^{\phi
}\mathcal{V}(\phi). \label{RayScalar-GBD}%
\end{equation}
As we can see, here again the (\ref{condition}) does not require the violation
of the strong energy condition (\ref{energy}). Thus, if we assume that
(\ref{energy}) holds, then any space-time described by these equations will
satisfy $R_{\mu\nu}V^{\mu}V^{\nu}\leq0$ , as long as
\begin{equation}
\omega(\phi)\dot{\phi}^{2}-e^{\phi}|V|^{2}\mathcal{V}(\phi)\geq0
\label{GBD-Cond}%
\end{equation}
We conclude that in this case the singularity theorem applies. On the other
hand, if (\ref{GBD-Cond}) is violated, then the solution may correspond to a
nonsingular space-time.\newline

In the following let us consider some solutions to (\ref{FE1-GBD},
\ref{FE2-GBD}), for some choices of the potential $\mathcal{V}(\phi)$ in the
case $\omega(\phi)=$ constant and $T_{\mu\nu}=0$. These solutions correspond
to homogeneous and isotropic models in GST theory and are obtained in the
Riemann frame $(M,\overline{g}=e^{-f}g,\overline{\phi}=0)$, in which case
$|V|^{2}=e^{\phi}$. If we take the line element written as in (\ref{FRW}),
then the field equations (\ref{FE1-GBD}) reduce to\newline%

\begin{equation}
3\frac{\dot{a}^{2}}{a^{2}}+3\frac{\epsilon}{a^{2}}=\frac{\omega}{2}\dot{\phi
}^{2}+\frac{e^{2\phi}}{2}\mathcal{V}(\phi), \label{L2}%
\end{equation}
\begin{equation}
2\frac{\ddot{a}}{a}+\left(  \frac{\dot{a}}{a}\right)  ^{2}+\frac{\epsilon
}{a^{2}}=-\frac{\omega}{2}\dot{\phi}^{2}+\frac{e^{2\phi}}{2}\mathcal{V}(\phi),
\label{L3}%
\end{equation}
while (\ref{FE2-GBD}) gives%
\begin{equation}
\ddot{\phi}+3\frac{\dot{a}}{a}\dot{\phi}=-\frac{e^{2\phi}}{\omega}\left(
\mathcal{V}(\phi)+\frac{1}{2}\frac{d\mathcal{V}}{d\phi}\right)  , \label{L4}%
\end{equation}
where $\epsilon=0,\pm1$, according to the curvature of the spatial
section\footnote{It is interesting to note here that from the above we can
obtain the following equation:
\begin{equation}
\dot{H}=-\frac{\omega}{2}\dot{\phi}^{2}+\frac{\epsilon}{a^{2}},
\end{equation}
This equation might be useful to set the possible values of $\omega$%
.}.\newline

Solutions to the equations (\ref{L2}), (\ref{L3}) and (\ref{L4}) with flat
spatial section $\epsilon=0$ were obtained for the following choices of the
scalar potential: $\mathcal{V}_{o}e^{-(2+\lambda)\phi}$, $e^{-2\phi}(m\phi
^{2}+\Lambda)$ and $2\lambda e^{-2\phi}(\phi^{2}+2\omega/3)^{2}$, where
$\mathcal{V}_{o}$, $\lambda$, $m$ and $\Lambda$ are constants. These, in the
Riemann frame, correspond to the dilaton field, the massive scalar field and a
field with quartic interaction. These three cases are displayed in the tables
below, where the presence of a singular behaviour is determined according to
whether or not (\ref{GBD-Cond}) holds:
\FloatBarrier
\begin{table}[th]%
\begin{tabular}
[c]{|c|c|c|}\hline
\rule[1ex]{0pt}{1ex} Potential & \rule[1ex]{0pt}{1ex} Solution to $\phi(t)$ &
\rule[1ex]{0pt}{1ex} Where\\\hline
\rule[0.5ex]{0pt}{2ex} $e^{-(\lambda+2)\phi}\mathcal{V}_{o}$ & \rule[0.5ex]%
{0pt}{2ex} $\phi(t) = \frac{2}{\lambda}\ln\left( \frac{H_{o}\lambda^{2}%
}{2\omega}t + e^{\frac{\lambda}{2}\phi_{o}} \right) $ & \rule[0.5ex]{0pt}{2ex}
$H_{o} = \pm\sqrt{\frac{\omega\mathcal{V}_{o}}{6\omega- \lambda^{2}}}$\\\hline
\rule[0.5ex]{0pt}{2ex} $e^{-2\phi}(m\phi^{2} +\Lambda)$ & \rule[0.5ex]%
{0pt}{2ex} $\phi(t) = \phi_{o} - \frac{2\alpha}{\omega}t$ & \rule[0.5ex]%
{0pt}{2ex} $\alpha= \pm\sqrt{-\frac{\omega\Lambda}{4}}$\\\hline
\rule[0.5ex]{0pt}{2ex} $e^{-2\phi} 2\lambda(\phi^{2} - 2/3\omega)^{2}$ &
\rule[0.5ex]{0pt}{2ex} $\phi(t) = \phi_{o}\exp\left( -\frac{4 A}{\omega
}t\right) $ & \rule[0.5ex]{0pt}{2ex} $A = \pm\sqrt{\frac{\lambda}{3}}$\\\hline
\end{tabular}
\caption{The other parameters $\lbrace\mathcal{V}_{o}, \lambda, m, \Lambda$
and $\phi_{o}\rbrace$ are constants.}%
\end{table}\begin{table}[th]%
\begin{tabular}
[c]{|c|c|c|}\hline
Potential & Solution to $a(t)$ & Singularity in finite time\\\hline
\rule[2ex]{0pt}{2ex} $e^{-(\lambda+2)\phi}\mathcal{V}_{o}$ & \rule[2ex]%
{0pt}{2ex} $a(t) = a_{o}\left[  \frac{\lambda^{2}H_{o}}{2\omega}
e^{-\frac{\lambda}{2}\phi_{o}}t + 1 \right] ^{2\omega/\lambda^{2}}$ & Singular
if $\omega> \lambda^{2}/6$\\\hline
\rule[2ex]{0pt}{2ex}$e^{-2\phi}(m\phi^{2} +\Lambda)$ & \rule[2ex]%
{0pt}{2ex}$a(t) = a_{o}\exp\left[  \alpha t\left( \phi_{o} -\frac{\alpha
}{\omega}t \right) \right] $ & Non-singular if $\omega< 0$\\\hline
\rule[2ex]{0pt}{2ex}$e^{-2\phi} 2\lambda\left( \phi^{2} - \frac{2}{3\omega
}\right) ^{2}$ & \rule[2ex]{0pt}{2ex}$a(t) = a_{o}\exp\left[ - \frac
{\omega\phi_{o}^{2}}{8}\exp\left( -\frac{8 A}{\omega}t-1\right)  +\frac
{2}{3\omega} A t\right] $ & Singular if $\omega> 0$\\\hline
\end{tabular}
\caption{The parameters are the same as in Table 1 and $a_{o}$ is also a
constant.}%
\end{table}\FloatBarrier

\section{Final remarks\label{sec6}}

The Hawking-Penrose singularity theorems are a direct consequence of
Einstein's theory of gravity. Given the important role they have played in our
understanding of the Universe, as modelled by general relativity, it is also
of interest to find out the analogous of these theorems in alternative gravity
theories. Singularity theorems and energy conditions have been studied in
connection with Brans-Dicke theory, perhaps the most popular scalar-tensor
gravity theories. However, their meaning remains still controversial due to
the question of whether or not the Einstein frame and the Jordan frame are
physically equivalent \cite{Faraoni1}. In the present geometrical approach,
this controversy does not arise as the physical entities defined in the theory
are naturally invariant under frame transformations \cite{Catena}.

\section*{Acknowledgments}

The authors thanks CNPq/CAPES for financial support. I. P. Lobo is supported
by the CAPES-ICRANet program (BEX 14632/13-6).

\end{document}